\documentclass[%
 reprint,
 amsmath,amssymb,
 aps,
 prl,nofootinbib
]{revtex4-2}
 \usepackage[hidelinks]{hyperref}
 \hypersetup{colorlinks=true,
     linkcolor=blue,
     filecolor=blue,
     urlcolor=blue,
     citecolor=blue,
     }
\usepackage{graphicx}
\usepackage{dcolumn}
\usepackage{bm}

\usepackage{kotex}
\usepackage{mathtools}
\usepackage{lipsum}
\usepackage{amsmath}
\newcommand{\RomanNumeralCaps}[1]
 {\MakeUppercase{\romannumeral #1}}             
\usepackage{mathrsfs}                          
\usepackage{upgreek}
\usepackage{bm}     

\begin{document}

\preprint{APS/123-QED}

\title{New Search for Dark Matter with Neutrons at Neutrino Detectors}

\author{Koun Choi}
\email{koun@ibs.re.kr}
\affiliation{Center for Underground Physics$,$ Institute for Basic Science$,$ Daejeon 34126$,$ Republic of Korea}

\author{Jong-Chul Park}
\email{jcpark@cnu.ac.kr}
\affiliation{Department of Physics and Institute of Quantum Systems (IQS)$,$ \\Chungnam National University$,$ Daejeon 34134$,$ Republic of Korea}

\begin{abstract}
Large-volume neutrino experiments are ideal for testing boosted dark matter (BDM) scenarios.
We propose, for the first time, an approach to utilize knockout neutrons by detecting de-excitation $\gamma$ rays and coincident captured neutrons from dark-matter interactions with bound neutrons in oxygen, while previous studies have focused on knockout-protons and electrons. 
This method is especially crucial for water \v{C}erenkov detectors, where high proton \v{C}erenkov threshold ($\sim$1 GeV) suppresses signal acceptance. 
Recently, Super-Kamiokande (SK) was doped with 0.03\% gadolinium (SK-Gd) to enhance neutron tagging efficiency.
Using SK-Gd as a target experiment, we demonstrate that this method increases sensitivity to BDM models by an order of magnitude compared to proton-based analysis, and it allows exploration of a wider range of light dark-matter models previously inaccessible with proton-based analysis. 
We also present the projected sensitivity for the upcoming Hyper-Kamiokande detector.
\end{abstract}

\maketitle
    
\noindent {\bf Introduction.} 
In the absence of WIMP signal, other well-motivated dark-matter (DM) models in the sub-GeV or even lower mass range are receiving rising attention~\cite{Arcadi:2017kky, Cooley:2022ufh}.
Thermal-relic DM with sub-GeV mass results in electron/nuclear recoils of energy below O(1) keV, which is challenging to detect. 
However, if DM is boosted by certain mechanisms originating from the nature of the dark sector~\cite{DEramo:2010keq, Belanger:2011ww, Agashe:2014yua, Kong:2014mia, Bhattacharya:2014yha, Kim:2016zjx}, astrophysical objects~\cite{Kouvaris:2015nsa, An:2017ojc, Calabrese:2021src, Wang:2021jic} or cosmic rays~\cite{Bringmann:2018cvk, Ema:2018bih, Cappiello:2019qsw, Jho:2021rmn}, it can be searched in large-volume DM detectors~\cite{Cherry:2015oca, Giudice:2017zke, COSINE-100:2018ged, Wang:2019jtk, Alhazmi:2020fju, Jho:2020sku, PandaX-II:2021kai, CDEX:2022fig, NEWSdm:2023qyb, COSINE-100:2023tcq, Guha:2024mjr, PandaX:2024pme, CDEX:2024qzq, Kim:2020ipj} or bigger volume detectors with MeV-scale
energy thresholds~\cite{Agashe:2014yua, Kong:2014mia, Necib:2016aez, Alhazmi:2016qcs, Kim:2016zjx, Chatterjee:2018mej, Kim:2018veo, Berger:2019ttc, Kim:2020ipj, DeRoeck:2020ntj, PROSPECT:2021awi, Granelli:2022ysi, Super-Kamiokande:2022ncz, Dutta:2024kuj}.

If the momentum transfer from DM is considerably large, single-nucleon interaction dominates and one can track the recoiled nucleon propagating in the medium. 
However, a water \v{C}erenkov detector (WCD) suffers from the high \v{C}erenkov momentum threshold of proton ($> 1.07$ GeV). 
Super-Kamiokande (SK) has searched for cosmic-ray boosted DM (BDM) in the limited signal momentum window (1.2 GeV $< p_p <$ 2.3 GeV) for free protons, and found no signal excess~\cite{Super-Kamiokande:2022ncz}.
One can extend this search to DM interaction with bound nucleons in oxygen. 

When DM scatters off a bound nucleon and ejects the nucleon from the nucleus, called `quasi-elastic' process:
\begin{equation}
\begin{array}{l}
\chi + ^{n}X \rightarrow \chi + ^{n-1}X + n (+ \gamma)\,, 
\\
\chi + ^{n}X \rightarrow \chi + ^{n-1}Y + p (+ \gamma)\,,
\end{array}
\end{equation}
the residual nucleus may de-excite by emitting $\gamma$ rays or other particles.
Thanks to sub-MeV energy resolution of liquid scintillator detectors such as Borexino or JUNO, almost background-free detection of de-excitation $\gamma$ rays is feasible~\cite{Cui:2022owf,Suliga:2023pve,Dutta:2024kuj}. 
In case of WCDs, the de-excitation $\gamma$ rays are buried in single electron-like events, making them hard to be detected. 
However, the knockout neutrons can be captured by ambient nuclei in the water and emit additional $\gamma$ rays with characteristic energy.
Search for timing and spatial coincidence of the de-excitation and neutron-capture $\gamma$ rays ($``\gamma + n"$ pair) offers an efficient background reduction, making searches for a rare signal promising in WCDs. 
Brief discussions on DM-oxygen quasi-elastic scattering have been in literature \cite{deNiverville:2016rqh,Cappiello:2019qsw} without detailed derivation of sensitivities to specific models.

In this work, we demonstrate a search for two-component BDM model~\cite{Belanger:2011ww, Agashe:2014yua} interacting with oxygen nucleus via a new vector boson coupled to the baryonic current, in realistic WCDs. 
We calculate the sensitivity of SK with 552.2-days data collected with 0.011\% mass concentration of gadolinium, and extrapolate the result to 10 years of Hyper-Kamiokande (HK) data taken with the same gadolinium concentration.

\medskip
\noindent {\bf Benchmark Scenario.} 
We consider a two-component DM scenario with a dark sector consisting of two complex scalars $\chi_0$ and $\chi_1$ with mass hierarchy $m_0 > m_1$, whose stability is protected by dark ${\rm U}(1)'\otimes{\rm U}(1)''$ gauge symmetries~\cite{Belanger:2011ww}. 
We assume that both $\chi_0$ and $\chi_1$ are charged under ${\rm U}(1)''$, while only $\chi_1$ is charged under ${\rm U}(1)'$. 
The dark sector is allowed to couple to the Standard Model (SM) sector only via ${\rm U}(1)'$.
The dark-sector gauge symmetries are assumed to be spontaneously broken leading to the dark gauge boson masses $m_{X'}$ and $m_{X''}$, respectively. 
Then, the lighter species $\chi_1$ directly interacts with the SM particles, whereas the heavier one $\chi_0$ does not.

As a benchmark example, we consider a baryophilic DM model, a simple and well-motivated extension of the SM, where the ${\rm U}(1)'$ gauge boson $X'$ couples to both $\chi_1$ and the SM quarks.
The relevant Lagrangian is
\begin{equation}
        \mathcal{L} \supset i q_B g_B X'_\mu [\chi_1^\dag \partial^\mu \chi_1 - (\partial^\mu \chi_1^\dag)\chi_1] + \frac{1}{3} g_B X'_\mu \bar q \gamma^\mu q\,,
    \label{eq:int}
\end{equation}
where $q_B$ is the baryon number of $\chi_1$ which is set to be 1, and $g_B$ is the ${\rm U}(1)'$ gauge coupling constant.
This sort of scenarios can arise with various new gauge bosons, e.g., ${\rm U(1)}_{B-L}$~\cite{Davidson:1978pm}, ${\rm U(1)}_{B}$~\cite{Nelson:1989fx}, $U(1)_{B-3L_{\mu,\tau}}$~\cite{Farzan:2016wym}, and
$U(1)_{T3R}$~\cite{Dutta:2019fxn}.
This model can be viewed as an effective field theory with a cutoff set well above the weak scale, staying irrelevant to the details of the UV completion that may occur above TeV~\cite{Batell:2014yra}.
We choose this model as it allows superior sensitivity with hadrons than leptons, also does not discriminate neutrons from protons in the interaction strength. 
We note that the search strategy with ``$\gamma + n$'' pair is valid for any new-physics scenario as far as there exist sufficient interaction with neutrons and substantial momentum transfer to kick out a neutron.

In the two-component DM model, while $\chi_0$ provides the dominant DM relic density and $\chi_1$ does the subdominant one, boosted $\chi_1$ particles produced through the $\chi_0 \chi_0^\dag \rightarrow \chi_1 \chi_1^\dag$ process today provide a good candidate for detection. 
Production of boosted $\chi_1$ occurs predominantly in the region of high $\chi_0$ density, in particular the Galactic Center. 
The resultant differential energy spectrum of $\chi_1$ is mono-energetic ($E_1=m_0$).
Integrating it over the entire sky for the NFW profile~\cite{Navarro:1995iw, Navarro:1996gj}, the flux of BDM $\chi_1$ can be estimated as
\begin{equation}
\Phi_1 \simeq 1.6 \times 10^{-4}\, \textnormal{cm}^{-2} \textnormal{s}^{-1} {\left(\frac{1\, \textnormal{GeV}}{m_0}\right)}^2\,,
\end{equation}
where we assume that the velocity-averaged
annihilation cross section $\langle \sigma_{\rm ann} v \rangle_{\chi_0 \chi_0^\dag \rightarrow \chi_1 \chi_1^\dag} = 5 \times 10^{-26}\, \textnormal{cm}^3/\textnormal{s}$ and the local dark-matter density near the Sun $\rho_\odot = 0.3~{\rm GeV}/{\rm cm}^{3}$~\cite{Kim:2018veo}.

\begin{figure}[t]
    \centering
    \includegraphics[width=0.5\textwidth]{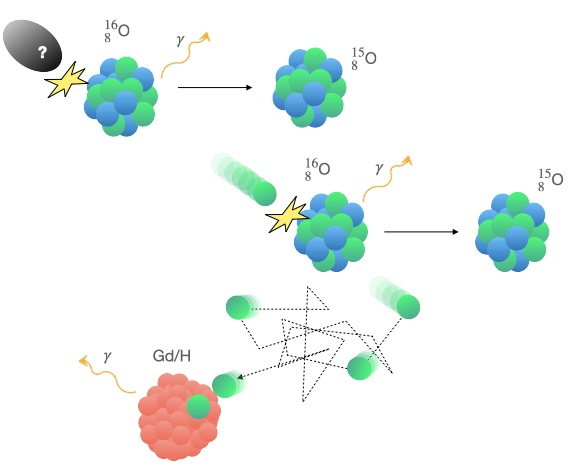}
    \caption{Schematic picture of a DM-oxygen scattering event.
    Green (blue) balls stand for neutrons (protons). 
    }
    \label{fig:drawing}
\end{figure}

Figure~\ref{fig:drawing} depicts the $\gamma$-ray production mechanism in three steps. Due to neutrons from final state interactions (FSIs) and secondary interactions, there is a substantial chance to detect the ``$\gamma + n$'' pair even in DM-proton collisions, but we conservatively consider only DM-neutron collisions.

\begin{figure}[t]
    \centering
    \includegraphics[width=0.53\textwidth]{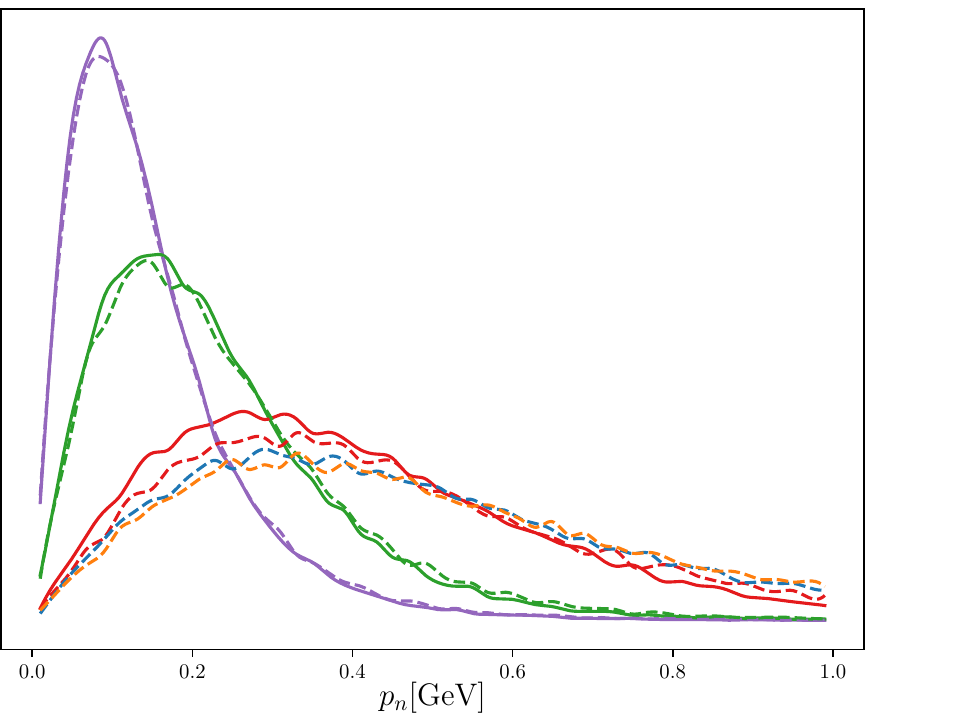}
    \caption{Momentum of kicked-out nucleons, $p_n$, for $m_0 =$ 1 GeV (solid) and 10 GeV (dashed), with $m_1 =$ 50 MeV (purple), 100 MeV (green), 500 MeV (red), 1 GeV (blue) and 5 GeV (orange), where the vector mediator mass $m_{X'} = 3 m_1$. The overall normalization is arbitrary.}
    \label{fig:pn}
\end{figure}

\medskip
\noindent {\bf Simulation of Dark Matter Signal.} 
When momentum transfer $Q^2 \gtrsim (\mathcal{O}(0.1\, \textnormal{GeV}))^2$, DM starts to resolve the inner structure of the nucleon.
The differential cross section of energetic DM-nucleon elastic scattering is given in Ref.~\cite{Batell:2014yra},
\begin{align}
&\frac{d \sigma_{1 N \rightarrow 1 N}} {dE_1^f} = \alpha_B q^2_B \label{eq:xsec}\\
&\times\frac{F^2_{1,N} A(E_1,E_1^f)+ F^2_{2,N} B(E_1,E_1^f)+F_{1,N} F_{2,N} C(E_1,E_1^f)}{(E_1^2-m_{1}^2) (m_{X'}^2 + 2m_N(E_1 -E_1^f)\}^2)}\,, \nonumber
\end{align}
where $\alpha_B = {g_B}^2/4\pi$.
The expressions for the form factors $F_{(1,2),N}$ and the kinematic functions $A, B, C$ are given in Appendix B of Ref.~\cite{Batell:2014yra}.
Figure~\ref{fig:pn} shows the distributions of the transferred momentum to nucleon from DM for the tested model parameter sets.
As seen in this plot, large fraction of recoiled nucleons in the tested parameter sets do not exceed the \v{C}erenkov momentum threshold of proton, which therefore highlights the importance of the ``$\gamma + n$'' signal search we propose.

\begin{figure}[t]
    \centering
    \includegraphics[width=0.53\textwidth]{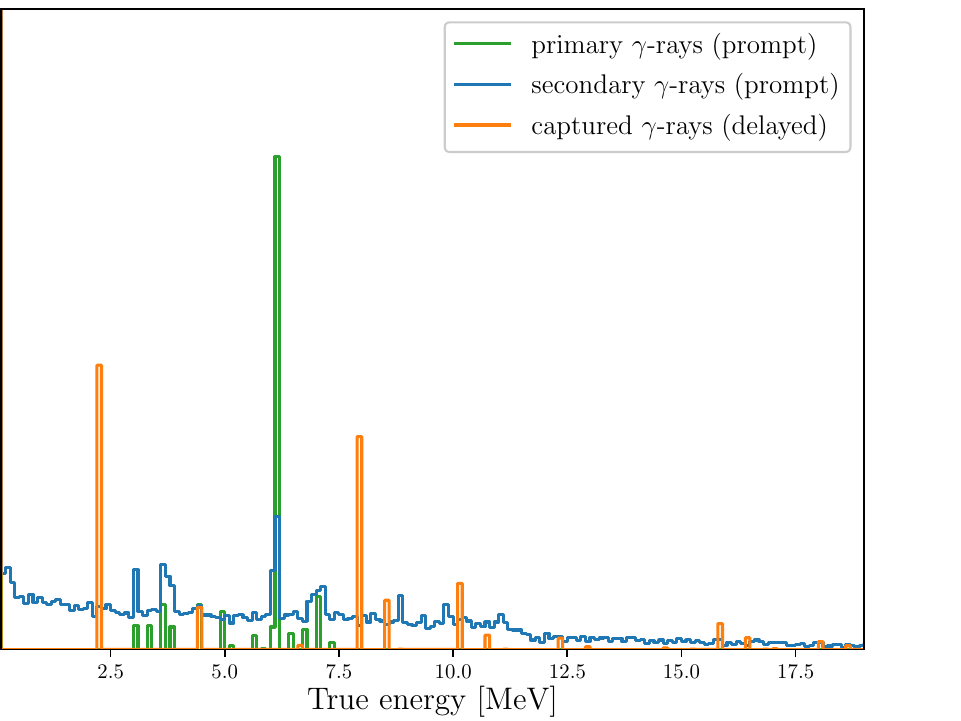}
    \caption{
    Summed energies of the primary (green) and the secondary $\gamma$ rays (blue, scaled by 10) per DM-oxygen interaction, and summed energy of the $\gamma$ rays per neutron capture by a hydrogen or a gadolinium (orange) for $m_0 =$ 10 GeV, $m_1 =$ 50 MeV and $m_{X'} = 3 m_1$. The overall normalization is arbitrary.
    }
    \label{fig:gammas}
\end{figure}

We then inject the momenta of the recoiled neutrons into \texttt{NEUT}~\cite{Hayato:2021heg}, a neutrino-interaction simulator with which de-excitation processes of residual nuclei can be simulated.
After DM knocks out a bound nucleon, whose momentum is assigned following a global fermi gas model, the nucleon experiences FSIs, resulting in the change of the number and spectra of outgoing particles.
The residual nucleus is excited with a nucleon hole state. 
The ``spectroscopic factor'', which can be regarded as a removal probability for each shell state, is calculated incorporating electron scattering measurements~\cite{Ankowski:2011ei,Benhar:2005dj}.
The remaining hole can be closed by de-excitation of the residual nucleus, which involves emission of nucleons or $\gamma$ rays. 
The $1p_{3/2}$ state de-excites by emitting 6.18 MeV $\gamma$ ray with a branching ratio of 86.9\% in the case of a neutron knock-out \cite{Abe:2014dyd,Leuschner:1994zz,Ajzenberg-Selove:1991rsl}, making the largest contribution to the primary $\gamma$ ray signal (green in Fig.~\ref{fig:gammas}).
As the $\gamma$-ray production probability is kept constant at sufficiently large momentum transfer~\cite{Ankowski:2011ei}, we choose model parameter sets with $Q^2 \gtrsim (40\, {\rm MeV})^2$ in most of the events, as can be seen in Fig.~\ref{fig:pn}.

The knockout nucleon, de-excitation $\gamma$ rays, and other particles are collected and propagated in the medium using \texttt{WCSim}~\cite{wcsim}, a \texttt{Geant4} based program for simulating WCDs.
The simulated volume is a cylindrical tank with a diameter of about 40 m and a height of 40 m, filled with gadolinium-doped water with mass concentration of 0.011\%.
During propagation, nucleons have a large chance to interact with other oxygen nuclei in water, emitting secondary particles such as nucleons and de-excitation $\gamma$ rays (blue in Fig.~\ref{fig:gammas}).
Following Ref.~\cite{Super-Kamiokande:2023oxd}, the \texttt{BERT} (\texttt{FTFP$\_$BERT$\_$HP} physics list) model~\cite{Wright:2015xia} is used to simulate these secondary interactions.
An observed prompt event has an energy sum from primary and secondary $\gamma$ rays.

Without gadolinium, neutrons are predominantly captured by hydrogen nuclei, emitting a single 2.2 MeV $\gamma$ ray. 
The typical neutron capture timing constant is a few hundreds $\mu$s, and thus $\gamma$ rays from neutron capture are collected in a separate timing window for a delayed signal attached to each prompt event trigger.
Techniques and algorithms for neutron tagging in water have been developed but suffer from the small number of photoelectrons from the 2.2 MeV $\gamma$ ray, resulting in the neutron tagging efficiency as low as $<25\%$. 
To increase the neutron tagging efficiency, the SK collaboration has loaded gadolinium corresponding to 0.011\% of the total mass in 2020~\cite{Super-Kamiokande:2021the}, a phase called SK-\RomanNumeralCaps{6} or SK-Gd. 
From the SK-\RomanNumeralCaps{6}, gadolinium takes up to 50\% the neutron capture processes as it has much higher cross section than hydrogen, resulting in a cascade of $\gamma$ rays which sum up to $\sim8$ MeV. 
The peaks from hydrogen and gadolinium captures can be clearly seen in Fig.~\ref{fig:gammas} (orange).
As a result, the detection efficiency of ``$\gamma + n$'' pair has been enhanced beyond 35\%~\cite{Super-Kamiokande:2023xup}.

\medskip
\noindent {\bf Background and Detection Efficiency.} 
The visible energy of a \v{C}erenkov ring seen by photomultiplier tubes (PMTs) is smeared by detector resolution which can be approximated with a gaussian function with a sigma value described in the Eq.~(2.3) of Ref.~\cite{Super-Kamiokande:2008ecj}, featuring $\sim20\%$ for 10 MeV electron.
After smearing the energy spectrum, a cut on the prompt signal 7.49 $< E <$ 29.49 MeV is applied, capturing 20--60\% of the signals of the tested models.
To estimate the realistic signal spectrum after background rejection including neutron tagging, we use the signal collection efficiency from Fig.~1 in Ref.~\cite{Super-Kamiokande:2023oxd}, calculated for each bin of prompt visible energy $E$ from 7.49 to 29.49 MeV with 2 MeV width.
For DM models with higher average momentum transfer than atmospheric neutrinos, larger neutron multiplicity is expected, which makes the neutrons easier to be paired with prompt signal.
In this case, applying the signal collection efficiency of the atmospheric neutrinos would be a conservative choice, and vice versa.

Figure~2 of Ref.~\cite{Super-Kamiokande:2023oxd} shows the remaining events after applying all reduction cuts to select neutral current quasi-elastic (NCQE) scattering atmospheric neutrino events. 
The majority of them are NCQE atmospheric neutrinos, but also other atmospheric neutrinos from NC non-QE or charged-current interactions. 
Around 5\% of the events come from non-atmospheric neutrinos such as spallation events, reactor neutrinos, and accidental coincidence events.
We take all of them as background events in our analysis.
For details of the background rejection cuts and neutron tagging technique, refer Refs.~\cite{Super-Kamiokande:2023oxd, Super-Kamiokande:2023xup}.

\begin{figure}[t]
  \begin{center}       
      \includegraphics[width=0.53\textwidth]{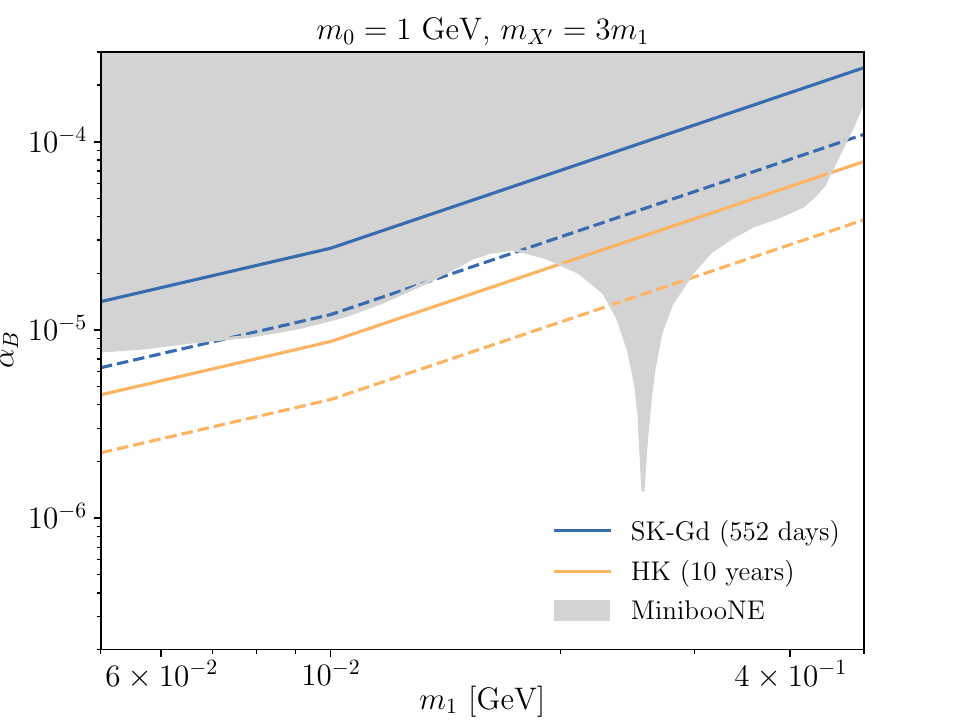}
      \includegraphics[width=0.53\textwidth]{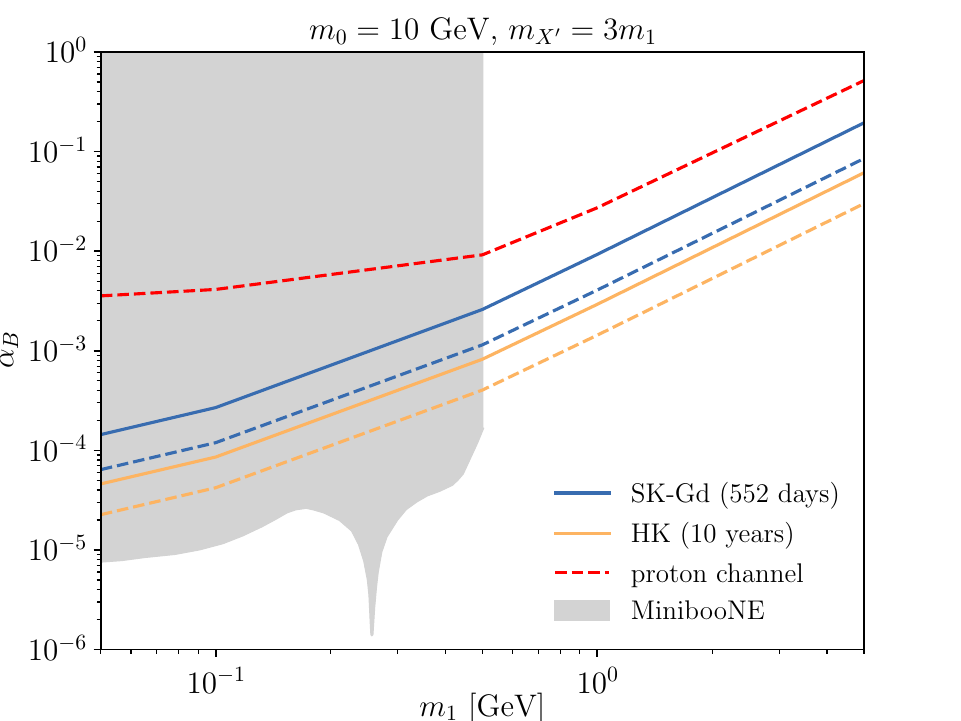}
      \caption{
  90\% C.L. sensitivity reaches on the coupling constant $\alpha_B$ as a function of lighter DM $\chi_1$ mass $m_1$ for SK-Gd 552.2 days (blue) and HK-Gd 10 years (orange). 
  The top (bottom) panel shows the results with heavier DM mass, $m_0=1$~GeV ($m_0=10$~GeV).
  Dashed lines indicate the limits before applying systematic uncertainties. 
  The gray shaded region displays the MiniBooNE bound~\cite{MiniBooNEDM:2018cxm}, which is independent from the choice of $m_0$. 
  In the bottom panel, the sensitivity for the SK-Gd 552.2 days data calculated using protons detected through their \v{C}erenkov light is also shown in red for comparison.
}
      \label{fig:eps}
  \end{center}
\end{figure}

\medskip
\noindent {\bf Sensitivity. } We now derive 552.2-days sensitivity to the baryophilic DM for the SK-Gd experiment.
We test the signal + background hypothesis against simulated background-only data by performing a binned likelihood analysis for 11 bins of visible energy.
The 90\% C.L. upper limit for DM-induced signal events is then converted to limit on $\alpha_B$.
In Fig.~\ref{fig:eps}, we display the sensitivity in the $m_1 - \alpha_B$ plane, assuming the heavier DM mass of $m_0 = 1$ or 10 GeV and the vector boson mass of $m_{X'} = 3m_1$.
We also extrapolate it to the expected sensitivity for 10 years of HK data assuming 0.011\% gadolinium concentration and the same detector resolution (orange).

In case of $m_0=10$ GeV, the momentum transfer in some collision can exceed the \v{C}erenkov threshold of proton. 
Therefore, we deliver the search sensitivity with free protons for the same SK-Gd data (552.2 days) for comparison, assuming negligible detector response and systematic uncertainties, which leads to a progressive constraint. 
We apply an angular cut $\cos\theta > 0$ to select forward-going events and a momentum cut $1.1 < p_p < 2.3$ GeV, and use the realistic signal collection efficiency and background taken from Ref.~\cite{Super-Kamiokande:2022ncz}. 

Across the entire DM mass range we tested, our method demonstrates substantially improved results compared to the proton-recoil detection counterpart, by a factor of six at the highest mass and larger at lower masses.
This can be explained by noticing that the event recoil spectrum stays sizable at lower momentum, so that absence of high \v{C}erenkov momentum cut in our method constitutes an advantage. 
Our method has stable signal acceptance not only for light DM scenarios but also for heavy DM scenarios, thanks to the property of de-excitation $\gamma$-ray being weakly tied to the recoiled nucleus energy (a caveat is that in case of finding a DM signal, we do not learn much about its property by analyzing the prompt signal only).

Our results are also compared with the existing constraint from the MiniBooNE beam-dump run~\cite{MiniBooNEDM:2018cxm} as an example (gray shaded), which however cannot be directly compared to ours as it does not originate from the two-component DM model~\footnote{Constraints on the two-component DM model from cosmological and astrophysical observations, depending on the details of the
dark sector, have been discussed in Refs.~\cite{Kamada:2021muh, Kim:2023onk}}.
Nevertheless, note that SK-Gd can probe a larger parameter space compared to a fixed-target experiment, as the available DM mass range of a fixed-target experiment is limited by the beam energy.

Theoretical prediction of the $\nu$-oxygen and DM-oxygen interaction suffers from large uncertainties.
The uncertainty in de-excitation $\gamma$-ray emission arises from the estimation of the spectroscopic factor and emission probabilities. Both prompt and delayed signals depend heavily on the choice of the secondary interaction model in detector simulation. The uncertainty in neutron tagging efficiency also affects the delayed signal. For a full set of systematic uncertainties, see Ref.~\cite{Super-Kamiokande:2023oxd}.
To take account in these uncertainties, we scale our upper limit by the ratio between the sizes of the statistical and total uncertainties found in Ref.~\cite{Super-Kamiokande:2023oxd}.

\medskip
\noindent {\bf Conclusions and Outlook. } 
This paper highlighted the unique sensitivity of large-volume water \v{C}erenkov detectors to BDM scenarios, which arises when the BDM candidate interacts with neutron such as in a gauged baryon number symmetry.
We showed that SK-Gd and HK data can test a wide range of currently unconstrained parameter space by searching for ``$\gamma + n$'' pairs.
In particular, they can explore lighter DM parameter space not accessible by recoiled proton signals, and scenarios with more than six times weaker couplings for the same masses.
Moreover, this approach has advantages over the fixed-target experiment in that the DM need not be lighter than the mediator particle produced by the beam and the explorable DM mass range is not limited by the beam energy.

Future searches in SK-Gd and HK could be improved in several aspects. 
First, gadolinium concentration has been further increased to 0.03\% since 2022~\cite{Abe:2024ydm}, which will permit higher neutron tagging efficiency.
Difference in the typical travel distances of neutrons from DM and neutrino interactions could be used to statistically separate them.
The large uncertainty in the interaction modeling can be improved by future nuclear experiments~\cite{Tano:2024uqa}. 
Improved modeling of atmospheric neutrino flux will reduce the background uncertainty.
Recently, the SK collaboration has announced the first observation of reactor neutrinos~\cite{reactor}, whose prompt positron signals typically fall below 8 MeV. 
The detection was possible by opening a neutron-search timing window using a lower-threshold trigger. 
Similar technique can be applied to lower the threshold of this analysis, and therefore to increase the signal acceptance to $\sim$6 MeV primary $\gamma$ rays.
The typical neutron travel distance in water is comparable to the vertex resolution of SK, making it difficult to reconstruct neutron direction for further background discrimination. 
However, future WCD like IWCD~\cite{nuPRISM:2014mzw, Hyper-Kamiokande:2018ofw} with high-granular multi-PMT (mPMT) sensors may enjoy some directional sensitivity. 
Detectors such as JUNO~\cite{JUNO:2021vlw} can also seek DM using the ``$\gamma + n$'' pair and neutron direction.

\medskip
\noindent {\bf Acknowledgments.}
We thank Yoshinari Hayato and Artur M. Ankowski for helpful conversations about oxygen de-excitation process. 
We are also grateful to Seiya Sakai for his help in analyzing NCQE signal.
The work of KC is supported by the fund from the Institute for Basic Science (IBS) under project code IBS-R016-Y2. 
The work of JCP is supported by the National Research Foundation of Korea grant funded by the Korea government(MSIT) (RS-2024-00356960).

\renewcommand{\refname}{References}
\bibliographystyle{utphys}
\bibliography{main}

\end{document}